\documentclass[twocolumn]{aastex631}

\usepackage{amsmath}
\usepackage{tabularx}

\begin{document}

\title{$S_8$ Tension in the Context of Dark Matter-Baryon Scattering}
 
\author{Adam He}
\affiliation{Department of Physics and Astronomy, University of Southern California, Los Angeles, CA 90089, USA}

\author{Mikhail M. Ivanov}
\affiliation{School of Natural Sciences, Institute for Advanced Study, 1 Einstein Drive, Princeton, NJ 08540, USA}
\affiliation{NASA Hubble Fellowship Program Einstein Postdoctoral Fellow}
\affiliation{Center for Theoretical Physics, Massachusetts Institute of Technology, Cambridge, MA 02139, USA}

\author{Rui An}
\affiliation{Department of Physics and Astronomy, University of Southern California, Los Angeles, CA 90089, USA}

\author{Vera Gluscevic}
\affiliation{Department of Physics and Astronomy, University of Southern California, Los Angeles, CA 90089, USA}

\begin{abstract}

We explore an interacting dark matter (IDM) model that allows for a fraction of dark matter (DM) to undergo velocity-independent scattering with baryons. In this scenario, structure on small scales is suppressed relative to the cold DM scenario.
Using the effective field theory of large-scale structure, we perform the first
systematic analysis of BOSS full-shape galaxy clustering data for the IDM scenario, and we find that this model ameliorates the $S_8$ tension between large-scale structure and \textit{Planck} data.
Adding the $S_8$ prior from DES to our analysis further
leads to a mild $\sim3\sigma$ preference for a non-vanishing DM-baryon scattering cross-section, assuming $\sim 10\%$ of DM is interacting
and has a particle mass of 1~MeV. This result produces a modest $\sim 20$\% suppression of the linear power at $k\lesssim 1~h$/Mpc, consistent with other small-scale structure observations.
Similar scale-dependent power suppression was previously shown to have the potential to resolve $S_8$ tension between cosmological data sets. The validity of the specific IDM model explored here will be critically tested with upcoming galaxy surveys at the interaction level needed to alleviate the $S_8$ tension.

\end{abstract}

\section{Introduction} \label{sec:intro}

An abundance of cosmological and astrophysical observations suggest that the majority of matter in the universe is non-baryonic 
dark matter (DM) (\cite{Bertone2004}). 
The current leading $\Lambda$CDM cosmological model posits that DM is cold and collisionless, 
and has held up well in light of cosmological data. 
However, there are tensions between cosmological parameters inferred from the early and the late universe probes under the $\Lambda$CDM model (\cite{Abdalla_2022}), which could be a consequence of unknown systematic errors (\cite{Bernal_2016, Di_Valentino_2021_S}) or an indication of new physics beyond $\Lambda$CDM (\cite{Di_Valentino_2021_H, Abdalla_2022}).

In particular, the $S_8$ tension at the 2.5$\sigma$-level is now established between large-scale structure (LSS) data and the cosmic microwave background (CMB) anisotropy measurements from \textit{Planck}, and recent studies have shown that scale-dependent suppression of the linear matter power spectrum might be able to resolve the $S_8$ tension. Such suppression can occur as a result of baryonic physics (\cite{Amon_2022}), or it may arise from new physics associated with dark energy and DM (\cite{Poulin_2022}). In this study, we consider a scenario where scale-dependent suppression of matter clustering occurs as a result of elastic collisions between a fraction of DM and baryons (\cite{Boddy2018}).
Such a scenario arises in compelling DM models, including the weakly interacting massive particles (WIMPs) and a whole landscape of new interacting DM (IDM) scenarios, and has been extensively studied and constrained with direct detection and cosmological probes (\cite{Snowmass_2013, battaglieri2017cosmic, akerib2022snowmass2021,Sigurdson2004,Dvorkin2014,Gluscevic_2018, Boddy_2018, Boddy2018, Boehm_2005, Xu_2021, Nguyen2021, Maamari_2021, Rogers_2022, Becker_2021, Nadler_2019, Nadler_2021, Li_2022, Gluscevic_2019, Slatyer_2018, Buen_Abad_2022, Hooper_2022}). 

In IDM cosmology, DM exchanges heat and momentum with baryons, and matter perturbations experience collisional damping, leading to a scale-dependent suppression of structure (\cite{B_hm_2001, Boehm_2005}) illustrated in Figure~\ref{fig:power spectra}. 
Galaxy clustering and lensing have not previously been used for parameter inference in IDM cosmology. At the same time, the best-fit value of $S_8$ shifts in the presence of IDM (\cite{Gluscevic_2018}), even when the linear cosmology is considered, including the CMB measurements from \textit{Planck} (\cite{Boddy2018, Nguyen2021}) and the Lyman-$\alpha$ forest data (\cite{Becker_2021, Hooper_2022, Rogers_2022}).
In this study, we derive the first bounds on the DM-baryon elastic scattering cross-section from galaxy clustering measured in the Baryon Oscillation Spectroscopic Survey (BOSS) (\cite{Alam_2017}), with and without the $S_8$ prior derived from the Dark Energy Survey data (\cite{Abbott_2022}).  
In particular, we consider elastic scattering between DM and protons\footnote{The correction arising from the presence of helium and other light elements is negligible, as shown in previous studies (\cite{Boddy2018}). We therefore use ``protons" and ``baryons" interchangeably here.}, and its effect on the $S_8$ tension between the early and late universe measurements of structure.

In addition to linear cosmology, the population statistic of the satellite galaxies in the Milky Way place stringent observational bounds on IDM (\cite{Nadler_2021}). Taking both Lyman-$\alpha$ forest measurements and Milky Way satellite measurements into account, the suppression of power in the range of scales corresponding to wavenumbers $0.2\lesssim k \lesssim2 \ h/\mathrm{Mpc}$ is only allowed up to 25\% (\cite{Chabanier_2019, Nadler_2021}). 
Interestingly, beyond-$\Lambda$CDM models
that alleviate $S_8$ tension tend to feature a specific form of the scale-dependent suppression in the linear transfer function. In particular, results in \cite{Amon_2022, Preston:2023uup} show a preference for a power-suppression plateau at small scales, inferred from a joint analysis of the weak lensing survey data and the CMB anisotropy. In the context of IDM, $\sim 10\%$ fractional cases also produce suppression of this form, while larger fractions gradually depart from the plateau feature (Figure~\ref{fig:fractions}).
For this reason, we focus on scenarios where $5-15\%$ of DM interacts with baryons, and undergoes collisional damping, while the rest of DM is collisionless. These cases are consistent with the bounds from the Milky Way satellite abundance and the Lyman-$\alpha$ forest data.
The linear matter power spectrum $P(k)$ and its non-linear corrections are illustrated in Figure~\ref{fig:power spectra}. 

In addition, direct detection constraints severely limit interactions for heavy DM particles, so we focus on sub-GeV DM candidates only (\cite{Kim:393744, Akerib_2017, Agnese_2018, Angle_2008, Aalseth_2013, Amole_2017, Angloher_2016, Angloher_2017, XENON:2018voc, PICO:2019vsc, CoGeNT:2012sne, SuperCDMS:2015eex, DAMIC:2019dcn, DarkSide:2018bpj}).
Our analysis considers the simplest scenario where velocity-independent and spin-independent scattering occurs between DM and protons; we leave a complete consideration of velocity-dependent scattering for future work. 

We find that the velocity-independent DM-baryon scattering with 10\% of DM allowed to scatter with baryons is consistent with both BOSS and \textit{Planck} data, and ameliorates the $S_8$ tension between LSS and CMB data. 
After combining BOSS and \textit{Planck} with weak lensing measurements from the Dark Energy Survey, there is a $\sim3\sigma$ preference for a non-zero interaction cross-section, at multiple DM particle masses. 
While the preference is mild, it is also consistent with all known observational bounds on DM interaction physics, and warrants further consideration. 
In particular, the preferred range of non-zero scattering cross-sections can be critically tested in the coming decade with a wide variety of small-scale structure probes, including the Lyman-$\alpha$ measurements from the Dark Energy Spectroscopic Instrument (DESI) (\cite{DESI:2016fyo}) and the census of dwarf galaxies from the Vera C.~Rubin Observatory (\cite{VeraRubin2018}).  
More generally, our results are indicative of the preference towards the scale-dependent power suppression that helps to reconcile cosmological data sets; this suppression of power is of similar nature to that seen in other proposed solutions to $S_8$ tension in the literature (\cite{Amon_2022, Ye_2021,Poulin_2022}).

This paper is organized as follows. In Section~\ref{sec:scattering}, we briefly describe the cosmology of IDM, and in Section~\ref{sec:methods}, we outline our methods. Section~\ref{sec:results} presents the key results of our data analysis. We discuss and conclude in Section~\ref{sec:discussion}.

\section{Matter Perturbations in IDM} \label{sec:scattering}

Within an IDM cosmology that features elastic scattering between DM and baryons, the linear Boltzmann equations contain interaction terms that capture momentum transfer between the two cosmological fluids (\cite{Gluscevic_2018, Boddy2018}),
\begin{equation}\label{boltzmann}
     \begin{split}
        \dot{\delta}_{\chi} = &-\theta_{\chi} - \frac{\dot{h}}{2}, \qquad \dot{\delta}_{\mathrm{b}} = -\theta_{\mathrm{b}} - \frac{\dot{h}}{2}, \\ \dot{\theta}_{\chi} = &-\frac{\dot{a}}{a}\theta_{\chi}+c^{2}_{\chi}k^{2}\delta_{\chi} + R_{\chi}\left(\theta_{\mathrm{b}}-\theta_{\chi}\right), \\
        \dot{\theta}_{\mathrm{b}} = &-\frac{\dot{a}}{a}\theta_{\mathrm{b}}+c^{2}_{\mathrm{b}}k^{2}\delta_{\mathrm{b}} + \frac{\rho_{\chi}}{\rho_{\mathrm{b}}}R_{\chi}\left(\theta_{\chi}-\theta_{\mathrm{b}}\right) \\ 
        &+ R_{\gamma}\left(\theta_{\gamma}-\theta_{\mathrm{b}}\right),
    \end{split}
\end{equation}
where subscripts ${\chi}$ and ${\mathrm{b}}$ denote DM and baryons, respectively; $\delta$ denotes density perturbations, $\theta$ represents velocity divergence; $h$ is the trace of the scalar metric perturbation; $c$ represents the sound speeds in respective fluids; $R_{\gamma}$ is the momentum transfer rate between baryons and photons from Compton scattering; and $R_{\chi}$ is the momentum transfer rate between DM and baryons from their non-gravitational interaction, 
\begin{equation}\label{Rchi}
    R_{\chi} = \frac{1}{3}\sqrt{\frac{2^7}{\pi}}
    \frac{a\rho_{\mathrm{b}}\sigma_{0}}{m_{\chi}+m_\mathrm{b}}\left(\frac{T_{\chi}}{m_{\chi}}+\frac{T_{\mathrm{b}}}{m_{\mathrm{b}}}+\frac{V^{2}_{\mathrm{RMS}}}{3}\right)^{-\frac{1}{2}},
\end{equation}
\noindent where $m_\chi$ is the DM particle mass, $m_\mathrm{b}$ is the mean baryon mass, and $T$ denotes fluid temperatures. The root-mean-square bulk relative velocity between DM and baryons is defined as (\cite{Dvorkin2014}),
\begin{equation}\label{RMS}
    V_{\mathrm{RMS}}^{2} = \left< \vec{V}_{\chi}^{2}\right>_{\xi} = \int \frac{dk}{k}\Delta_{\xi} \left(\frac{\theta_{\mathrm{b}}-\theta_{\chi}}{k^{2}}\right)^{2},
\end{equation}
\noindent where $\Delta_{\xi}$ is the primordial curvature variance per log wavenumber $k$. 
An integral over $k$ appearing in the Boltzmann equations introduces mode mixing; however, an analytic approximation for the bulk relative velocity that remains constant for $z > 10^{3}$ and scales linearly with $z$ for $z\lesssim 10^{3}$ is used to reproduce the effect of $V_{\mathrm{RMS}}^{2}$ on $R_{\chi}$ with a precision adequate for cosmological analyses (\cite{Tseliakhovich2010, Dvorkin2014, Gluscevic_2018, Boddy2018}).\footnote{This model of relative bulk velocity is an approximation, and we make this choice for concreteness only; the value of relative bulk velocity does not have an observable effect in the context of velocity-independent DM scattering.}
To solve the Boltzmann equations in the presence of IDM, we use a modified version of the Boltzmann solver \texttt{CLASS}, which allows for DM-baryon scattering parameterized by a momentum transfer cross-section $\sigma_0$ (\cite{Gluscevic_2018, Boddy2018}).\footnote{We note that this model corresponds to the power-law parametrization of the momentum transfer cross section, $\sigma_\mathrm{MT}=\sigma_0 v^n$, for $n=0$, previously used in the literature (\cite{Gluscevic_2018, Boddy2018, Dvorkin2014}).}

In order to make a prediction for late-time evolution of the matter power spectrum on scales corresponding to galaxy clustering, weak lensing, and related LSS observables, we merge the modified IDM \texttt{CLASS} code with a \texttt{CLASS-PT} module, previously developed as a tool for the computation of LSS power spectra in the mildly non-linear regime (\cite{Baumann:2010tm,Carrasco_2012,Cabass:2022avo,Ivanov:2022mrd}).\footnote{\url{https://github.com/Michalychforever/CLASS-PT}}
\texttt{CLASS-PT} is a non-linear perturbation theory extension of \texttt{CLASS} that calculates non-linear 1-loop corrections to the linear matter power spectrum, and outputs the redshift-space galaxy power spectrum (\cite{Chudaykin2020}). 

\begin{figure}[t!]
\includegraphics[scale=0.53]{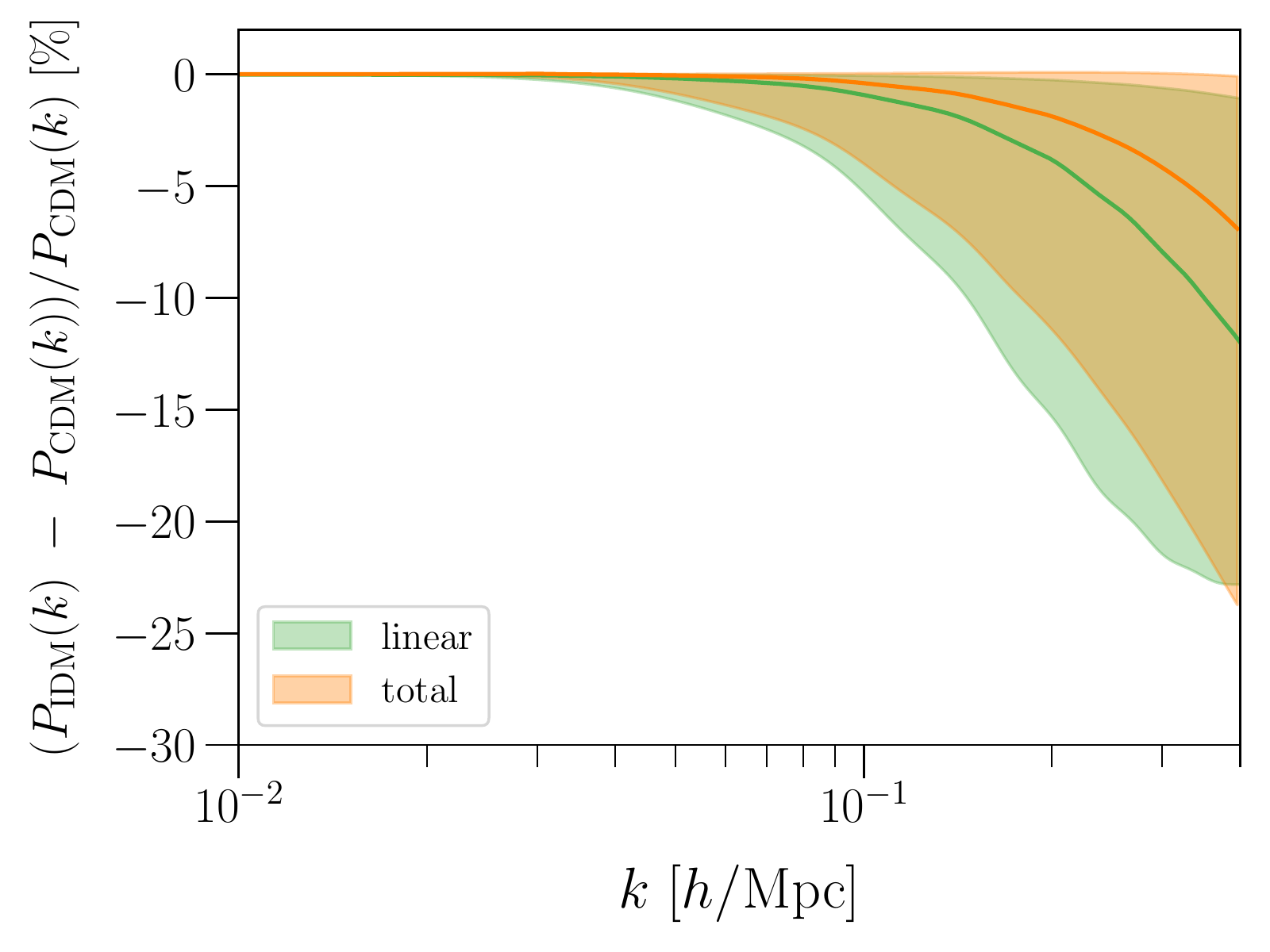}
\caption
{Percent difference between the matter power spectrum for an IDM cosmology with DM-baryon scattering and collisionless CDM cosmology. The linear power spectrum is shown in green, and the total power spectrum is shown in orange. The lines are generated with the best-fit parameter values from a joint \textit{Planck} + BOSS + DES analysis of the IDM model with a DM mass of 1 MeV, 
and interacting fraction $f_\chi=10\%$. The shaded bands designate the uncertainty in reconstructed matter power spectrum that corresponds to a 1$\sigma$ uncertainty around these best-fit parameter values. An increase in the interaction cross-section and in the interacting fraction lead to a greater suppression in $P(k)$, while the former also shifts the onset of suppression to larger scales. 
\label{fig:power spectra}}
\end{figure}

The formalism implemented in \texttt{CLASS-PT} rests on the effective theory of LSS, which should, in principle, be modified in the presence of non-gravitational interactions
between baryons and DM.
However, in the case of velocity-independent interaction, DM-baryon scattering only affects the evolution of matter perturbations at very high redshifts, where non-linear effects are entirely negligible. 
At all redshifts relevant to galaxy surveys, the DM-baryon interactions are effectively frozen, and the evolution of structure proceeds as in $\Lambda$CDM, with a suppressed initial power spectrum, shown in Figure~\ref{fig:power spectra}.\footnote{We show that DM-baryon interactions only impact
the evolution of matter perturbations at redshifts before recombination in Appendix~\ref{sec:structure evolution}.}
This means the standard implementation of \texttt{CLASS-PT} is entirely applicable to predicting late-time LSS observables in our scenario of interest.

\section{Data and analysis methods} \label{sec:methods}

We analyze the full \textit{Planck} 2018 TT, TE, EE, and lensing power spectra (\cite{Planck2018_V}), along with anisotropic galaxy clustering data from BOSS DR12 at $z=0.38$ and 0.61 (\cite{Alam_2017,Ivanov_2020,Ivanov:2019hqk}, see also~\cite{Zhang:2021yna,Chen:2021wdi}). As in~\cite{Chudaykin:2020ghx,Philcox:2021kcw}, our analysis is performed up to $k_\mathrm{max} = $ 0.2~$h/\mathrm{Mpc}$ for the galaxy power spectrum multipoles, 
from $ 0.2<k<  0.4$~$h/\mathrm{Mpc}$ for the real-space
power spectrum proxy $Q_0$~\citep{Ivanov:2021fbu}, and up to 
$k_\mathrm{max} = $ 0.08~$h/\mathrm{Mpc}$
for the bispectrum monopole~\citep{Ivanov:2021kcd,Philcox:2021kcw}.\footnote{Our BOSS full-shape likelihood for \texttt{CLASS-PT} is publicly 
available at~\mbox{\url{https://github.com/oliverphilcox/full_shape_likelihoods}}. }
We also add the post-reconstructed BOSS DR12 BAO data 
to this dataset following~\cite{Philcox:2020vvt}.
We stress that our EFT-based 
full-shape analysis 
is quite conservative as we consistently 
marginalize over all necessary 
nuisance parameters that 
capture galaxy bias, baryonic feedback, 
non-linear redshift space-distortions, 
etc. \cite{Philcox:2021kcw}.\footnote{Note that the priors used in our likelihood
are significantly wider than
the ones chosen in the EFT-based full-shape analysis of~\cite{Zhang:2021yna}.
Our choice ensures that our main
cosmological 
results are independent of priors 
on nuisance parameters. Note also that
in contrast to~\cite{Zhang:2021yna}, our priors are motivated by the physics
of BOSS red luminous galaxies, 
see~\cite{Chudaykin:2020ghx} for more detail.
}
Thus, our analysis is agnostic 
about the details of galaxy formation.
Note that we fit the BOSS galaxy 
clustering data within the 
IDM scenario in a fully consistent and rigorous manner, without any hidden $\Lambda$CDM assumptions. 
This can be contrasted with the 
standard ``compressed'' BOSS likelihood consisting of the BAO and RSD parameters that are derived assuming a
fixed \textit{Planck}-like $\Lambda$CDM template for
the underlying linear matter power spectrum~\citep{Ivanov_2020,eBOSS:2020yzd}. Moreover, our EFT-based 
likelihood includes the galaxy power 
spectrum 
shape information 
that is missing in the standard BOSS
likelihood (\cite{Alam_2017}),
see \cite{Ivanov_2020}
for a detailed discussion. 

\begin{figure}[t!]
\includegraphics[scale=0.53]{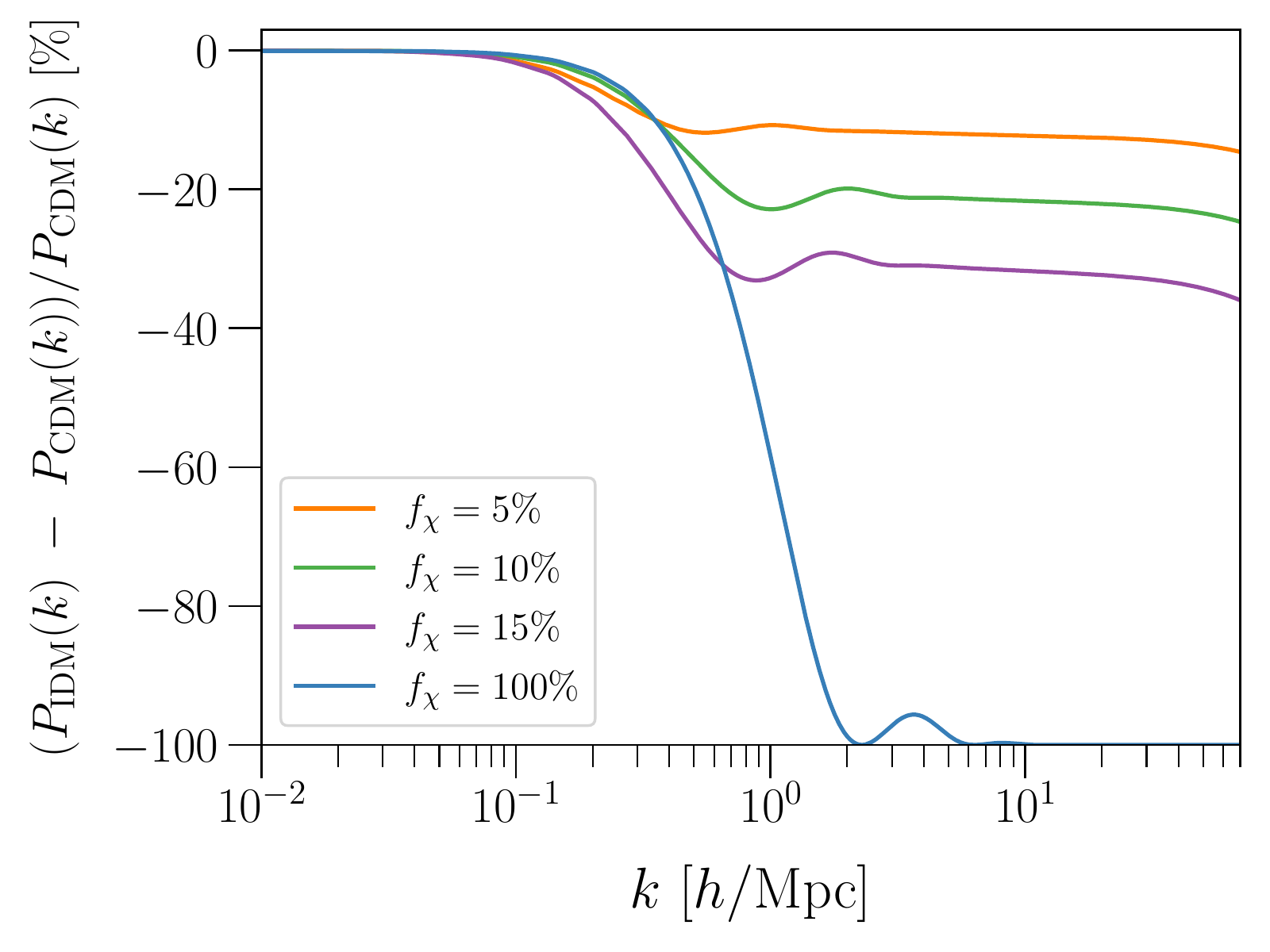}
\caption{Effect of varying the interacting DM fraction $f_\chi$ on the linear matter power spectrum for a cosmology with DM-baryon scattering. The residuals of the linear spectra w.r.t. CDM are shown for $f_\chi = 5\%$, 10\%, 15\%, and 100\%. These spectra are generated with the best-fit parameter values from a joint \textit{Planck} + BOSS + DES analysis of each fraction for a DM mass of $m_\chi=1$ MeV. Fractions higher than $\sim10\%$ generate a strong power suppression in the range $0.2\lesssim k \lesssim2 \ h/\mathrm{Mpc}$, for this choice of cross section, and are 
thus ruled out by Lyman-$\alpha$ forest and Milky Way satellite data.}
\label{fig:fractions}
\end{figure}

We also include weak lensing data from the Dark Energy Survey (DES-Y3), and we argue that the full DES-Y3 likelihood for DM-baryon interactions can be captured with a simple prior: $S_8$ = 0.776 $\pm$ 0.017. This is because $S_8$ is measured by DES to be the same value for $\Lambda$CDM, WDM, and $\Lambda$CDM extensions like early dark energy (EDE), indicating a robustness under different cosmological models (\cite{Abbott_2022, DES_extensions_2021, Hill_2020,Ivanov:2020ril}), as long as the late-time growth of structure is not modified in these models. Also, $S_8$ is close to being model independent as it is the primary directly 
observed principle component 
of the weak lensing data. 
In future work, the full DES-Y3 likelihood for DM-baryon scattering should be calculated to confirm our argument; in the meantime, we will treat a prior on $S_8$ as equivalent to adding the complete DES-Y3 dataset to our analysis.

We use the \textit{Planck}, BOSS, and DES data sets as proxies to data that drive the $S_8$ tension. We do not consider weak lensing from KiDS-1000 or HSC-Y3 data because their joint analysis with DES-Y3 necessitates modeling of the full covariance (\cite{DES_2022}), which is beyond the scope of analyses done in this work. We do not include Pantheon+ data (\cite{Brout_2022}) because this data would only further constrain $\Omega_\mathrm{m}$; since $\Omega_\mathrm{m}$ is not appreciably correlated with $\sigma_0$, we do not expect this data to alter our results in a significant way.
We have checked that the inclusion of eBOSS DR16 BAO does not strenghten our constraints either. Work is underway to convert eBOSS DR16 into a full-shape likelihood, and to incorporate Lyman-$\alpha$ data into our analysis (see e.g.~(\cite{Ivanov:2021zmi,Chudaykin:2022nru})). 
Preliminary 
reports on the suppression on the growth 
in these data sets~(e.g.~\cite{Ivanov:2021zmi,Goldstein:2023gnw}; and references therein) suggest that they may further favor IDM, but future studies are necessary to confirm this expectation.

To obtain bounds on our IDM model for parameters of interest, such as $H_0$, $S_8$, and the momentum transfercross-section $\sigma_0$, we perform Markov Chain Monte Carlo (MCMC) parameter estimation using our merged \texttt{CLASS} code. 
We use the MCMC sampler \texttt{MontePython} and interface it with our version of \texttt{CLASS} (\cite{Brinckmann_2018, Audren_2012}). 
We choose the Metropolis-Hastings algorithm and assume flat priors on $\{\omega_{\mathrm{b}}$, $\omega_{\mathrm{DM}}$, 100$\theta_\mathrm{s}$, $\tau_\mathrm{reio}$, $\mathrm{ln}(10^{10}A_{\mathrm{s}})$, $n_{\mathrm{s}}$$\} + \sigma_{0}$.\footnote{Following related studies of $S_8$ tension, we chose a flat prior on $\sigma_0$ and we discuss prior dependence of our results in Appendix~\ref{sec:log prior}. As expected, a log-flat prior still results in a mild preference for interactions.}
Following \cite{Gluscevic_2018}, we fix the IDM particle mass $m_\chi$ in each MCMC fit and consider the following benchmark particle masses: 100 keV, 1 MeV, 20 MeV, and 100 MeV. We choose this mass range because of the strict constraints on IDM from direct detection above 1 GeV, and constraints on $N_\mathrm{eff}$ that rule out masses lower than $\sim1$ MeV (\cite{Lewin1995}, \cite{An_2022}). 
We set the fraction of DM that interacts with baryons $f_\chi$ to be 10\%, while the rest of the DM behaves as CDM; then, we perturb the parameter space slightly and explore fractions $f_\chi=5\%$, 7.5\%, 12.5\%, and 15\%. 
We model free-streaming neutrinos as two massless species and one massive species for which $m_\nu$ = 0.06 eV, in line with the \textit{Planck} convention (\cite{Planck2018_VI}). 
A chain is deemed converged if the Gelman-Rubin convergence criterion $|R-1|$ is less than 0.01.

\section{Results} \label{sec:results}

\begin{figure}[t!]
\includegraphics[scale=0.275]{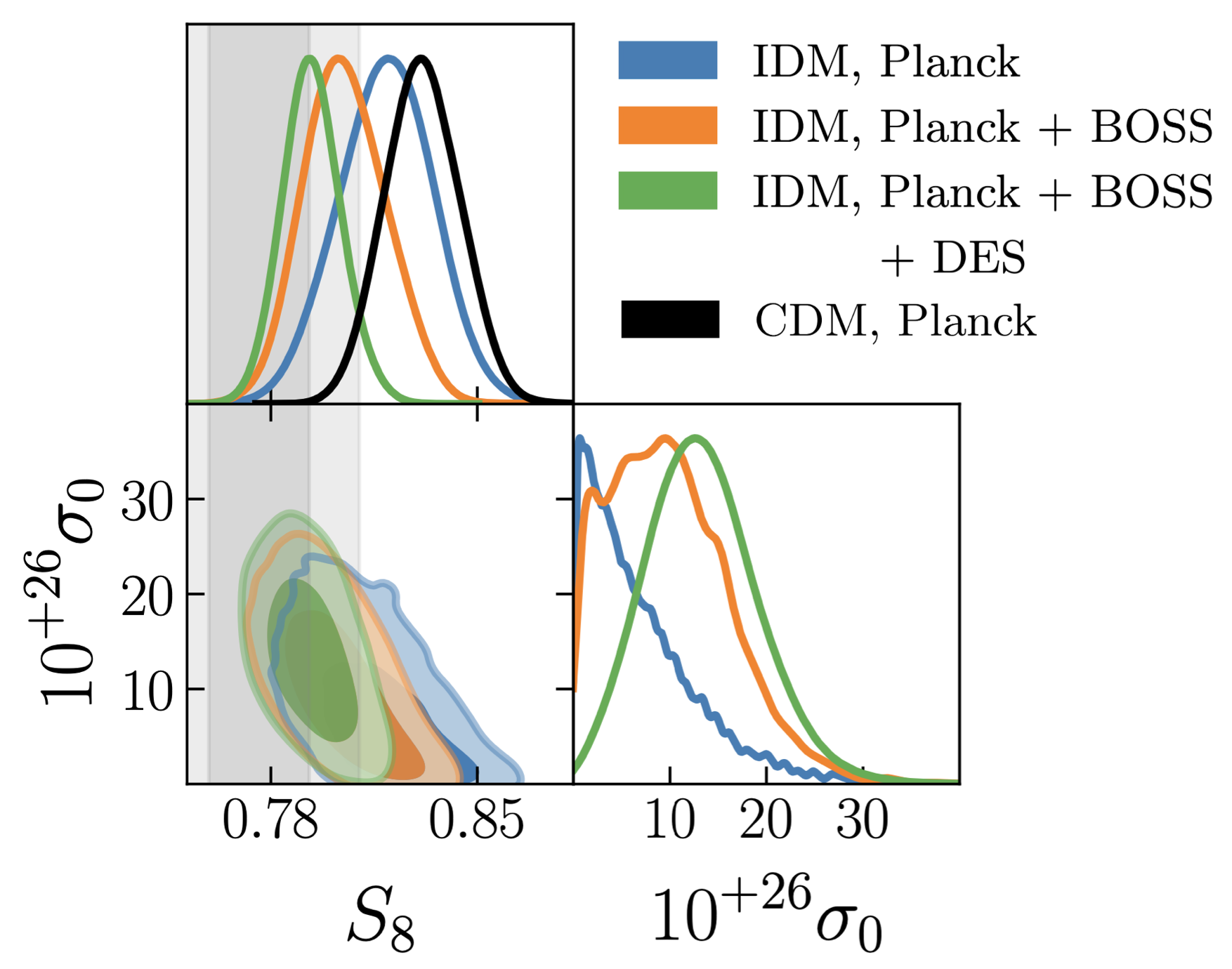}
\caption{68\% and 95\% confidence-level marginalized posterior distributions of relevant parameters for $\Lambda$CDM from \textit{Planck} (black) and our fractional DM-baryon interacting model (colored) from different combinations of \textit{Planck}, BOSS, and DES data. The gray bands show the DES measurement of $S_8$. The bottom right-hand panel shows a 2.6$\sigma$ preference for non-zero interactions between DM and baryons under a combined \textit{Planck}, BOSS, and DES analysis.
\label{fig:small triangle plot}}
\end{figure}

We find that for all masses tested in the range $[0.1,100]$~MeV, our model ameliorates the standard $S_8$ tension between \textit{Planck} and DES by 30\%, decreasing it from 2.6$\sigma$ to 1.8$\sigma$, while leaving the $H_0$ tension unchanged. 
When \textit{Planck} is combined with BOSS, our model reduces the $S_8$ tension with DES 
from 2.6$\sigma$ to 1.3$\sigma$. Since the 
qualitative picture 
is the same for all 
models considered, we
choose the $m_\chi=1$~MeV case
as a baseline. Our results for the $m_\chi=1$~MeV, $f_\chi=10\%$ model are shown in Figure~\ref{fig:small triangle plot}, which displays 1D and 2D marginalized posterior distributions for relevant parameters in our analysis, compared to standard results under $\Lambda$CDM. We note that our IDM model does not impact $\Omega_\mathrm{m}$; our model only affects perturbations and not background quantities, which is why $\Omega_m$ is indistinguishable from its value in $\Lambda$CDM.

Note the strong degeneracy between $\sigma_0$ and $S_8$. This degeneracy is unsurprising: a higher cross-section means that DM and baryons are more strongly coupled, leading to a greater suppression in the power spectrum and a corresponding decrease in $S_8$. Therefore, imposing a prior on $S_8$ that favors lower values leads to a mild preference for larger cross-sections. Indeed, we observe this in Figure~\ref{fig:small triangle plot}: BOSS data prefer a lower value of $S_8$ as compared to \textit{Planck} (\cite{Philcox:2021kcw}), so the combination of \textit{Planck} and BOSS shifts the $\sigma_0$ posterior towards larger cross-sections (shown in orange). The combination of \textit{Planck} and BOSS data with the $S_8$ prior from DES further disfavors low cross-sections, resulting in a preference for non-zero interactions between DM and baryons (shown in green). Overall, Figure~\ref{fig:small triangle plot} shows that the marginalized posterior from a combined \textit{Planck} + BOSS + DES analysis is maximized for $\sigma_0$ value of $1.32^{+0.52}_{-0.65}\times 10^{-25} \ \mathrm{cm}^2$ (at $68\%$ CL).

\begin{table*}
\centering
\tabletypesize{\scriptsize}
\tablewidth{0pt} 
\caption{The maximum of the marginalized posterior (the maximum of the full posterior) and ± 68\% confidence level uncertainties for cosmological parameters of interest under a \textit{Planck} + BOSS + DES analysis of our fractional IDM model with a flat prior on $\sigma_0$, compared to $\Lambda$CDM. The last row shows the difference in $\chi^2$ with respect to $\Lambda$CDM. All IDM values are for $m_\chi$ = 1 MeV, $f_\chi$ = 10\%.} \label{tab:chi_flat}
\begin{tabular}{|c|c|c|c|}
\hline
Model & $\Lambda$CDM, \textit{Planck} + BOSS + DES & IDM, \textit{Planck} + BOSS + DES \\
\hline 
\hline
$\sigma_0 \ \mathrm{[}10^{-26} \  \mathrm{cm^2]}$& -- & $13.23 \ (5.163)^{+5.2}_{-6.5}$ \\ [0.5ex] 

$\Omega_\mathrm{m}$ & $ 0.308 \ (0.307)\pm{0.005}$ & $0.311 \ (0.309)\pm{0.005}$ \\ [0.5ex] 

$\sigma_8$ & $ 0.802 \ (0.806)\pm{0.005}$ & $0.780 \ (0.792)_{-0.009}^{+0.007}$ \\ [0.5ex] 

$S_8$ & $ 0.813 \ (0.815)\pm{0.009}$ & $0.794 \ (0.804)^{+0.009}_{-0.01}$ \\ [0.5ex] 
\hline
$\Delta\chi^2_{\mathrm{min}}$ & -- & $-6.7$ \\
[0.5ex]
 \hline
\end{tabular}
\end{table*}

In Table~\ref{tab:chi_flat}, we show the values of the relevant parameters that maximize the marginalized posterior and the full posterior (the latter referred to as the best-fit), obtained from a \textit{Planck} + BOSS + DES analysis of the $f_\chi=10\%$, $m_\chi=1$~MeV model, as well as the $\chi^2$ statistics. We present the full list of constraints on all cosmological parameters for this scenario in Appendix~\ref{sec:constraints}, and for completeness we show full posterior distributions for this model in Appendix~\ref{sec:posteriors}. The $f_\chi=10\%$, $m_\chi=1$~MeV IDM model and $\Lambda$CDM present similar $\chi^2$ values when analyzed under \textit{Planck} only; however, we find $\Delta\chi^2$ = $-3.48$ once we include BOSS data, and $\Delta\chi^2$ = $-6.7$ once we include the DES prior on $S_8$, corresponding to a $2.6\sigma$ preference for non-zero interactions. We note that the preference for non-zero 
DM-baryon interactions is present even 
in the BOSS data alone: the IDM fit
that contains a single additional free parameter reduces $\chi^2$ by $3.02$, compared to the CDM model. We find that our fractional IDM model shows a consistent preference over CDM, regardless of the DM interacting fraction tested; this is discussed further in Appendix~\ref{sec:chisq}.

\section{Summary and Discussion} \label{sec:discussion}

We considered the concordance of cosmological data in the presence of velocity-independent scattering of baryons with sub-GeV DM particles in the early universe, for scenarios where the interacting component constitutes only a fraction of the total DM abundance. We found that this model is consistent with both the BOSS and \textit{Planck} data, and ameliorates the $S_8$ tension between LSS and the CMB. After combining BOSS and \textit{Planck} with the DES weak--lensing prior, we find a $2.6\sigma$ preference for non-zero interaction cross-section, for a range of DM particle masses and for an interacting fraction $\sim 10\%$.  

Our results have implications for DM searches and cosmology in general. Importantly, the model for DM interactions we considered here is quite broad and encompasses a number of well-motivated UV-complete scenarios where DM scatters with normal matter through a heavy mediator exchange at low energies, similar to the WIMP-like scenarios sought in direct detection (\cite{Gluscevic_2014, Snowmass_2013}). In fact, the scattering interactions that would trigger direct detection (at DM masses above 1 GeV) would also lead to momentum transfer with cosmological consequences, explored here for sub-GeV particles. This study is thus directly complementary to direct detection, as it explores different DM mass and cross-section regimes. 

The preference we find for non-zero interaction cross-section when combining data from \textit{Planck}, BOSS, and DES implies a preference for scale-dependent suppression in the linear matter power spectrum, similar to that seen in other beyond-CDM models that alleviate the $S_8$ tension (\cite{Poulin_2022, Ye_2021, Amon_2022}). Our expectation is that scale dependence is the primary driver of this preference, for the following reasons. First, the CMB and LSS in general probe different scales~(\cite{Amon_2022,Rogers:2023ezo}); we note that the CMB lensing 
has contributions from low redshifts ($z<1$), and yet its measurements of $S_8$ are consistent with those derived from the primary CMB anisotropy~(\cite{Planck_VIII},\cite{qu2023atacama}), consistent with our expectation.
Furthermore, interacting DM only affects matter perturbations long before recombination (\cite{Gluscevic_2018}), acting to ameliorate the tension through scale dependence of the linear power spectrum, rather than modifying clustering of matter at later times. 
Indeed, Refs.~\cite{Amon_2022} and \cite{Preston:2023uup} showed that the tension can be framed as a preference for a specific $k$-dependence of the transfer function, which happens to match the fractional IDM case, as discussed in our work; see Figure \ref{fig:fractions}. 

While the preference we find is mild, and could be a result of a random statistical fluctuation, or of an unknown systematic effect (\cite{ChavezMontero2022, Amonetal_2022, Leauthaud:2016jdb}), the specific model we consider here has several qualities that set it apart from models that were considered previously in the same context. First, it does not exacerbate  the $H_0$ tension, which is a common drawback of many models proposed to resolve the $S_8$ tension (\cite{Abdalla_2022}). Furthermore, it is simple and generic, relying on DM interaction physics that was proposed independent of the status of cosmological concordance. 

On the other hand, we wish to point out that the interpretation of the $S_8$ tension in the context of IDM has the same caveat as most beyond-$\Lambda$CDM models considered for the same purpose, in that it features more degrees of freedom than a vanilla cosmology. Namely, this study focused on exploring DM interactions that produce power suppression of the form preferred by the combination of cosmological data which alleviate the $S_8$ tension; we thus only varied the interaction cross section, while setting  $f_\chi$ and $m_\chi$ to fixed values. While concrete UV-complete models for DM may not allow the freedom in choosing the fraction and the particle mass, for the effective description of the interaction and for cosmological purposes, these two parameters could be treated as additional degrees of freedom of a larger class of models that produce velocity-independent scattering at low energies. Within that context, a full model-selection exercise should be performed to assess whether data favor IDM or a vanilla cosmology. However, this analysis exceeds the scope of the present work. 

The preferred range of interaction cross-sections and the corresponding level of power suppression in the linear matter power spectrum we consider here is consistent with all current observations (\cite{Nadler_2019,Nadler_2021,Maamari_2021, Nadler_2019, Nadler_2021, Xu_2018, Rogers_2022, Becker_2021, Hooper_2022}). Interestingly, the census of the MW satellite galaxies allows for up to $\sim 25-30\%$ decrement in power at $k\sim 30 \ h/\mathrm{Mpc}$, as compared to CDM (\cite{Nadler_2019,Nadler_2021}); this means that small-scale structure probes are on the verge of being able to detect the interacting DM signal necessary to resolve the tension, likely to be achieved within the next decade. 
We note that even the combined analyses of the existing measurements from eBOSS (\cite{eBOSS:2020yzd}), KiDS (\cite{KiDS:2020suj}), and HSC (\cite{HSC:2018mrq}) may be able to put further pressure on this model.  
Beyond the existing data, the matter power spectrum on quasi-linear and non-linear scales will be measured at high precision with upcoming surveys from DESI, Euclid~(\cite{Chudaykin:2019ock,Sailer:2021yzm}), and the Vera C. Rubin Observatory (\cite{Nadler_2019,drlicawagner2019}). 
In addition, Stage-3 and Stage-4 data that target high resolution measurements of the CMB, including SPT-3G (\cite{SPT-3G:2014dbx}), ACT (\cite{ACT:2020gnv}), the Simons Observatory (\cite{SimonsObservatory:2018koc}), CMB-S4 (\cite{CMB-S4:2016ple}), and CMB-HD (\cite{CMB-HD:2022bsz}) will probe quasi-linear scales, providing a longer lever arm for testing details of the scale dependence in the linear matter power spectrum. This study and the recent 
analyses of the $S_8$ tension (e.g.~\cite{Amon_2022,Poulin_2022,Rogers:2023ezo,Preston:2023uup}) highlight the need for further explore predictive models that affect the distribution of matter in the universe.

\textit{Acknowledgements.}~We thank the anonymous referee for their suggestions and careful review of our paper. We gratefully acknowledge the support from explore.org and Sylvie and David Shapiro at USC. We thank Ethan O.~Nadler for providing valuable comments on the manuscript. 
The work of MMI has been supported by NASA through the NASA Hubble Fellowship grant \#HST-HF2-51483.001-A awarded by the Space Telescope Science Institute, which is operated by the Association of Universities for Research in Astronomy, Incorporated, under NASA contract NAS5-26555. RA and VG acknowledge the support from NASA through the Astrophysics Theory Program, Award Number 21-ATP21-0135. VG acknowledges the support from the National Science Foundation (NSF) under Grant No. PHY-2013951. 

\pagebreak

\clearpage

\appendix
\section{Evolution of structure at low redshifts}
\label{sec:structure evolution}

\begin{figure*}[ht!]
\includegraphics[scale=0.65]{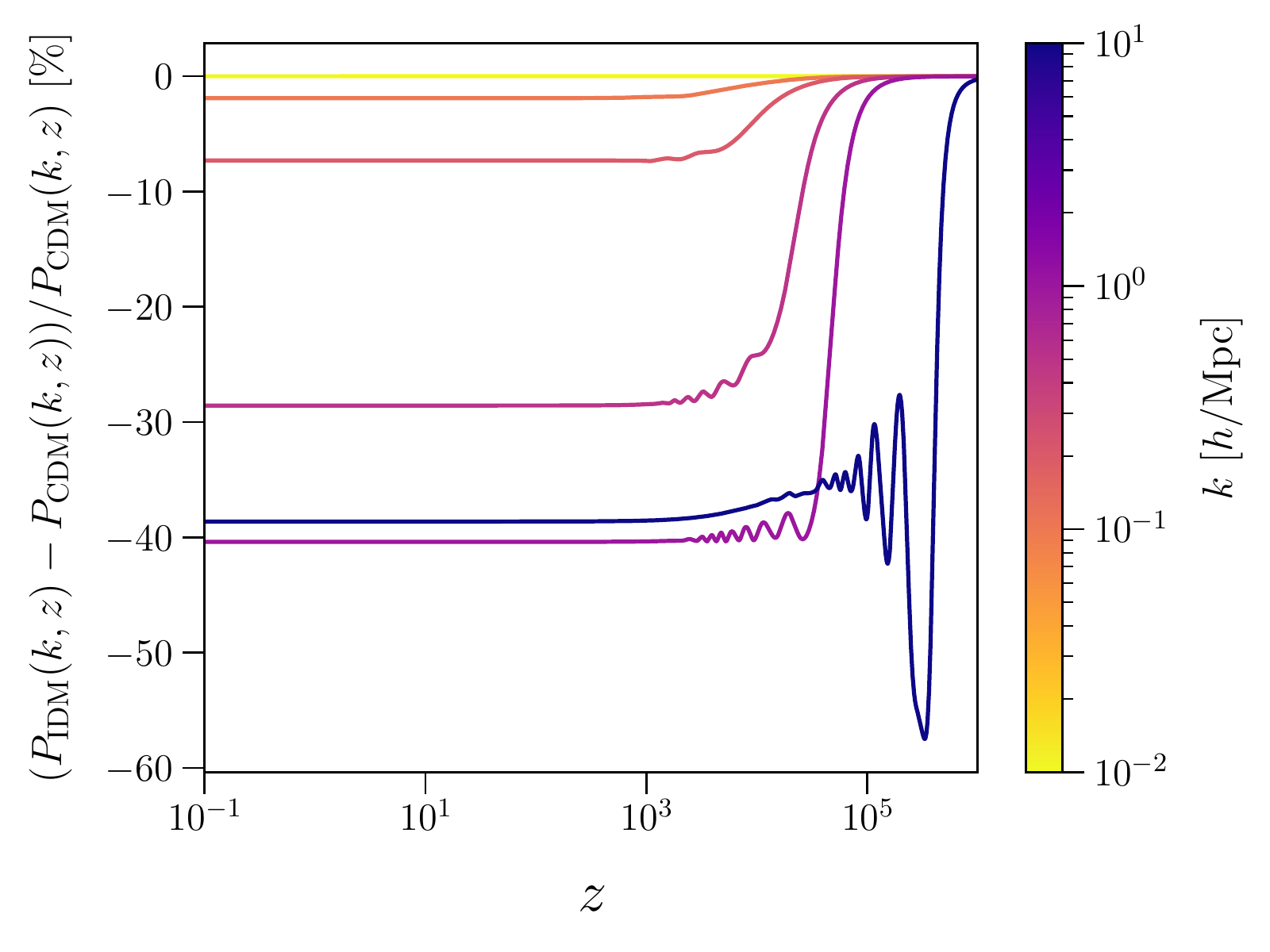}
\centering
\caption{Residual between the power spectrum for our fractional IDM model and the power spectrum for $\Lambda$CDM as a function of redshift, for different values of $k$. None of the curves continue oscillating past $z \sim 10^3$, indicating that there is no evolution on these scales past recombination. This means that DM-baryon interactions do not have an effect at these redshifts; if they did, the code would need to be edited to account for them. This plot is generated assuming best-fit cosmological parameters from a \textit{Planck} + BOSS + DES analysis of the $f_\chi=10\%$ IDM model, with a DM particle mass $m_\chi=1$ MeV and cross-section $\sigma_0=5.16\cdot10^{-26} \ \mathrm{cm}^2$.
\label{fig:transfer}}
\end{figure*}

We verify that there are no alternations in structure evolution on \textit{any} wavelength modes after recombination for our fractional IDM model by plotting the residual of the IDM power spectra with respect to $\Lambda$CDM as a function of redshift for different $k$ (Figure~\ref{fig:transfer}). We may also take the linear power spectrum generated for DM-baryon interactions and feed it into the standard non-linear CDM pipeline implemented by \texttt{CLASS-PT}, without introducing any additional counterterms to the non-linear power spectrum calculation.

\clearpage

\section{Full cosmological parameter constraints}
\label{sec:constraints}

We display the full set of cosmological parameter constraints that correspond with the minimum $\chi^2$ value in a \textit{Planck} + BOSS + DES analysis of the $f_\chi=10\%$, $m_\chi = 1$ MeV model in Table~\ref{tab:constraints}.

\begin{table*}[htb!]
\tabletypesize{\scriptsize}
\tablewidth{0pt} 
\centering
\caption{Full parameter constraints for a \textit{Planck} + BOSS + DES analysis of the  $f_\chi=10\%$, $m_\chi = 1$ MeV model. The top half displays bounds for standard cosmological parameters, and the bottom half displays bounds on EFT bias parameters. ``Best fit'' refers to the maximum of the full posterior, while ``Marginalized max'' refer to the maxima of the marginalized posteriors.
The upper scripts (1), (2), (3), (4) 
of 
the galaxy bias parameters
$b_1,b_2,b_{\mathcal{G}_2}$
refer to the NGC $z=0.61$, SGC $z=0.61$,
NGC $z=0.38$, SGC $z=0.38$ BOSS DR12 data chunks,
respectively. 
} 
\label{tab:constraints}
\begin{tabular}{|l|c|c|c|c|}
 \hline
Parameter & Best-fit & Marginalized max $\pm \ \sigma$ & 95\% lower & 95\% upper \\ \hline
$100~\omega{}_\mathrm{b }$ &$2.246$ & $2.254_{-0.015}^{+0.014}$ & $2.226$ & $2.283$ \\
$\omega{}_\mathrm{DM }$ &$0.1192$ & $0.12_{-0.001}^{+0.00098}$ & $0.118$ & $0.1219$ \\
$100~\theta{}_\mathrm{s }$ &$1.042$ & $1.043_{-0.00054}^{+0.00045}$ & $1.042$ & $1.044$ \\
$\mathrm{ln}(10^{10}A_\mathrm{s })$ &$3.04$ & $3.042_{-0.015}^{+0.015}$ & $3.012$ & $3.072$ \\
$n_\mathrm{s }$ &$0.9698$ & $0.9725_{-0.0055}^{+0.0045}$ & $0.9628$ & $0.9827$ \\
$\tau{}_\mathrm{reio }$ &$0.05215$ & $0.05288_{-0.0072}^{+0.0072}$ & $0.03864$ & $0.06727$ \\
$10^{+26}\sigma{}_0$ &$5.163$ & $13.23_{-6.5}^{+5.2}$ & $1.55$ & $24.57$ \\
$z_\mathrm{reio }$ &$7.44$ & $7.491_{-0.7}^{+0.76}$ & $6.048$ & $8.956$ \\
$\Omega{}_{\Lambda }$ &$0.691$ & $0.6888_{-0.0052}^{+0.0054}$ & $0.6786$ & $0.6992$ \\
$Y_\mathrm{He}$ &$0.2479$ & $0.2479_{-6.1e-05}^{+6e-05}$ & $0.2478$ & $0.248$ \\
$H_0$ &$67.88$ & $67.84_{-0.39}^{+0.39}$ & $67.1$ & $68.6$ \\
$10^{+9}A_\mathrm{s }$ &$2.09$ & $2.096_{-0.031}^{+0.032}$ & $2.033$ & $2.158$ \\
$\sigma_8$ &$0.7921$ & $0.7796_{-0.0094}^{+0.0068}$ & $0.764$ & $0.7967$ \\
\hline
$b^{(1)}_{1 }$ &$2.029$ & $2.013_{-0.044}^{+0.046}$ & $1.923$ & $2.102$ \\
$b^{(1)}_{2 }$ &$-0.6918$ & $-0.5745_{-0.6}^{+0.54}$ & $-1.679$ & $0.558$ \\
$b^{(1)}_{{\mathcal{G}_2} }$ &$-0.5745$ & $-0.4302_{-0.29}^{+0.28}$ & $-0.9879$ & $0.1287$ \\
$b^{(2)}_{1 }$ &$2.145$ & $2.167_{-0.056}^{+0.056}$ & $2.058$ & $2.274$ \\
$b^{(2)}_{2 }$ &$-0.1907$ & $-0.3824_{-0.67}^{+0.61}$ & $-1.611$ & $0.883$ \\
$b^{(2)}_{{\mathcal{G}_2} }$ &$-0.1638$ & $-0.1909_{-0.34}^{+0.33}$ & $-0.8453$ & $0.4748$ \\
$b^{(3)}_{1 }$ &$1.941$ & $1.929_{-0.043}^{+0.043}$ & $1.845$ & $2.014$ \\
$b^{(3)}_{2 }$ &$0.0399$ & $-0.1347_{-0.52}^{+0.48}$ & $-1.095$ & $0.8526$ \\
$b^{(3)}_{{\mathcal{G}_2} }$ & $-0.367$ & $-0.3617_{-0.28}^{+0.28}$ & $-0.9123$ & $0.184$ \\
$b^{(4)}_{1 }$ &$1.96$ & $1.966_{-0.055}^{+0.056}$ & $1.858$ & $2.074$ \\
$b^{(4)}_{2 }$ &$-0.6163$ & $-0.4016_{-0.58}^{+0.53}$ & $-1.481$ & $0.7131$ \\
$b^{(4)}_{{\mathcal{G}_2} }$ &$-0.4559$ & $-0.3989_{-0.32}^{+0.32}$ & $-1.029$ & $0.231$ \\
\hline
\end{tabular}
\end{table*}

\clearpage

\section{$\chi^2$ statistics}
\label{sec:chisq}

Table~\ref{tab:models} shows $\chi^2$ values for all masses and data combinations tested in our analysis of the $f_\chi=10\%$ IDM model.
We also test different fractions under a \textit{Planck} + BOSS + DES analysis to see whether or not our finding is specific to the $f_\chi=10\%$ model; Table~\ref{tab:[pbd]} shows the $\chi^2$ statistics from these runs. We test $f_\chi=5\%$, $7.5\%$, $12.5\%$,
and $15\%$; Figure~\ref{fig:fractions} shows the effect that changing $f_\chi$ has on the linear matter power spectra of our IDM model. From Table~\ref{tab:[pbd]}, it is clear that our fractional IDM model shows a consistent preference over CDM, regardless of DM interacting fraction.

\raggedbottom

\begin{table*}[b]
\tabletypesize{\scriptsize}
\tablewidth{0pt} 
\caption{$\Delta\chi_\mathrm{min}^2$ values for different masses and data set combinations tested in our analysis, for an interacting fraction $f_\chi=10\%$. Each $\Delta\chi_\mathrm{min}^2$ value is given with respect to the CDM $\chi^2$ value for that data set.} 
\label{tab:models}
\begin{tabular}{|c||c|c|c|c|c|}
\hline
$m_\chi$ & \textit{Planck} & \textit{Planck} + BOSS & \textit{Planck} + BOSS + DES & BOSS + DES & BOSS \\
\hline 
\hline
{100 keV}& +2.48 & $-3.34$ & $-4.78$ & $-0.416$ & $-2.398$ \\ [0.5ex] 
{1 MeV}& +1.08 & $-3.48$ & $-6.7$ & $-0.248$ & $-3.02$ \\ [0.5ex] 
{20 MeV}& +1.42 & $-5.26$ & $-6.42$ & $-0.734$ & $-1.996$ \\ [0.5ex] 
{100 MeV}& +3.42 & $-3.3$ & $-3.72$ & +0.034 & $-2.56$ \\ [0.5ex]
\hline
\end{tabular}
\end{table*}

\raggedbottom

\begin{table*}[htb]
\tabletypesize{\scriptsize}
\tablewidth{0pt} 
\centering
\caption{$\Delta\chi_\mathrm{min}^2$ values of different fractions and masses for our IDM model under a \textit{Planck} + BOSS + DES analysis. Each $\Delta\chi_\mathrm{min}^2$ value is given with respect to the CDM $\chi^2$ value for this data set combination.}
\label{tab:[pbd]}
\begin{tabular}{|c||c|c|c|c|c|}
\hline
$m_\chi$ & $f_\chi=5\%$ & $f_\chi=7.5\%$ & $f_\chi=10\%$ & $f_\chi=12.5\%$ &
$f_\chi=15\%$ \\
\hline 
\hline
100 keV & $-3.66$ & $-4.8$ & $-4.78$ & $-3.04$ & $-4.54$ \\ [0.5ex] 
1 MeV & $-2.42$ & $-5.08$ & $-6.7$ & $-4.2$ & $-4.84$ \\ [0.5ex] 
20 MeV & $-2.38$ & $-5.28$ & $-6.42$ & $-5.44$ & $-5.44$ \\ [0.5ex] 
100 MeV & $-2.74$ & $-2.88$ & $-3.72$ & $-3.88$ & $-3.88$ \\ [0.5ex]
\hline
\end{tabular}
\end{table*}

\pagebreak

\section{Posteriors}
\label{sec:posteriors}

\begin{figure*}[ht!]
\includegraphics[scale=0.65]{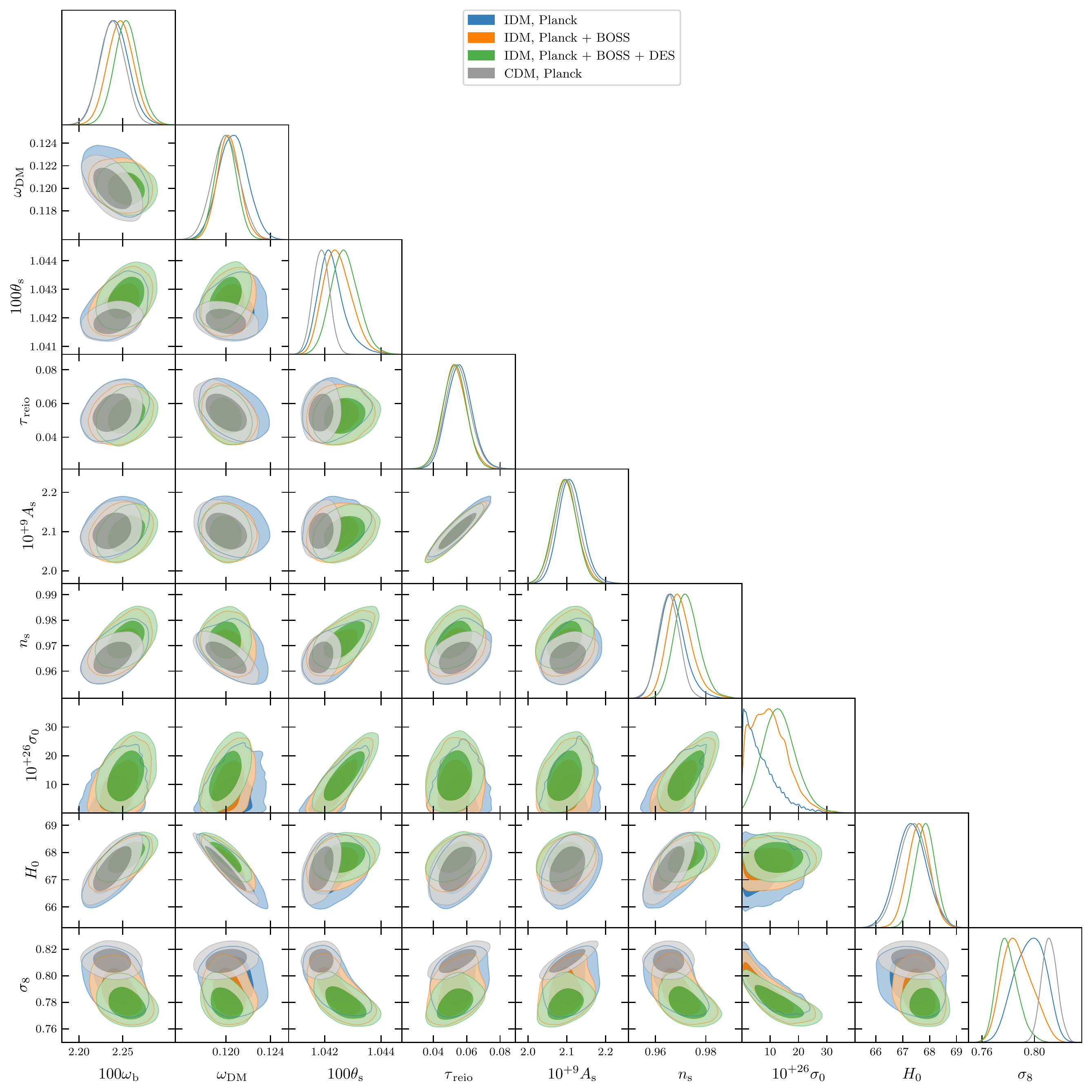}
\caption{68\% and 95\% confidence level marginalized posterior distributions of all cosmological parameters for $\Lambda$CDM from \textit{Planck} (gray) and our $m_\chi=1$ MeV, $f_\chi=10\%$ DM-baryon interacting model (colored) from different combinations of \textit{Planck}, BOSS, and DES data.
\label{fig:triangle plot}}
\end{figure*}

\begin{figure*}[ht!]
\includegraphics[scale=0.65]{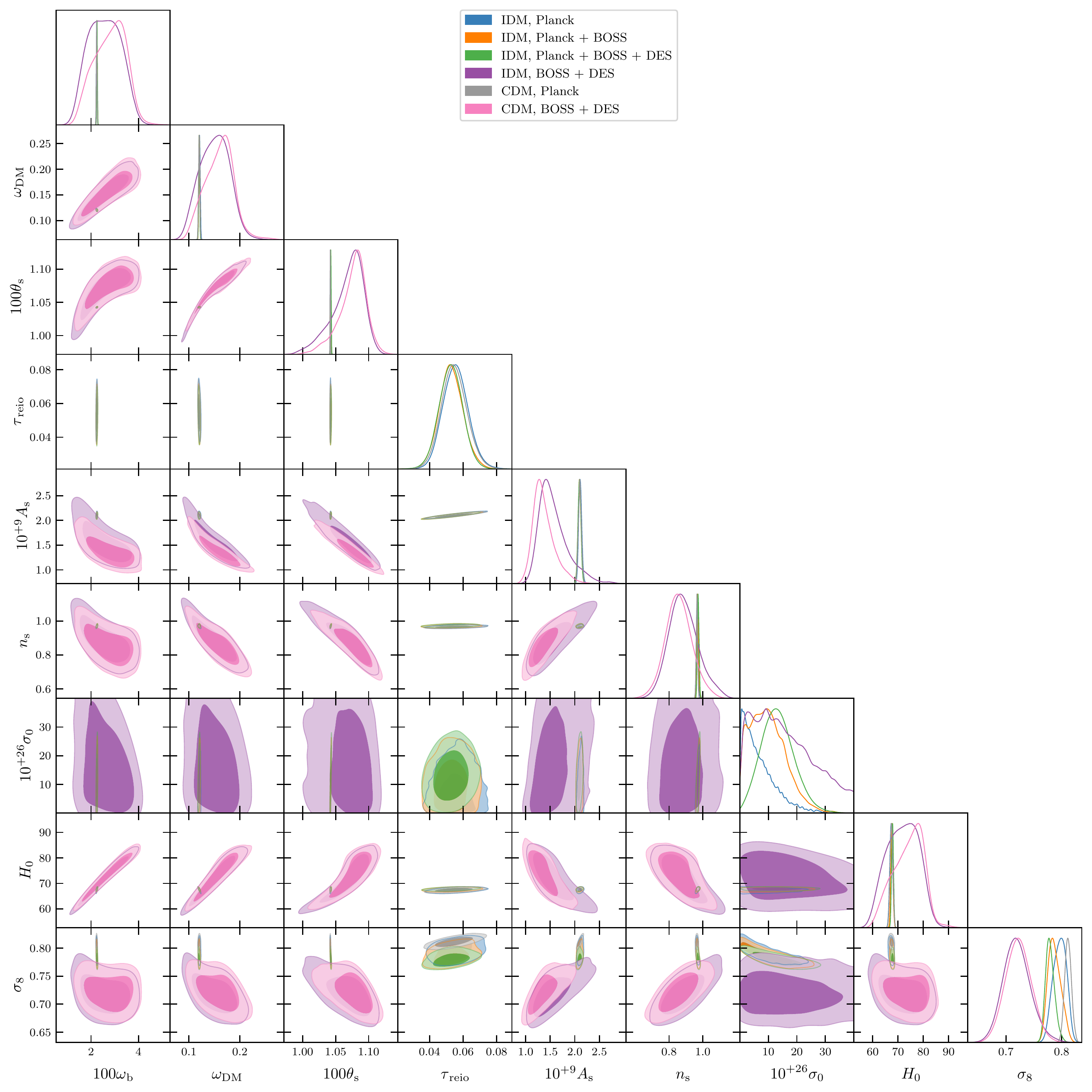}
\caption{68\% and 95\% confidence level marginalized posterior distributions of all cosmological parameters for $\Lambda$CDM from \textit{Planck} (gray) and our $m_\chi=1$ MeV, $f_\chi=10\%$ DM-baryon interacting model (colored) from different combinations of \textit{Planck}, BOSS, and DES data. Same as Figure~\ref{fig:triangle plot}, but with posteriors for BOSS + DES added.
\label{fig:BOSS DES triangle plot}}
\end{figure*}

\begin{figure*}[ht!]
\includegraphics[scale=0.65]{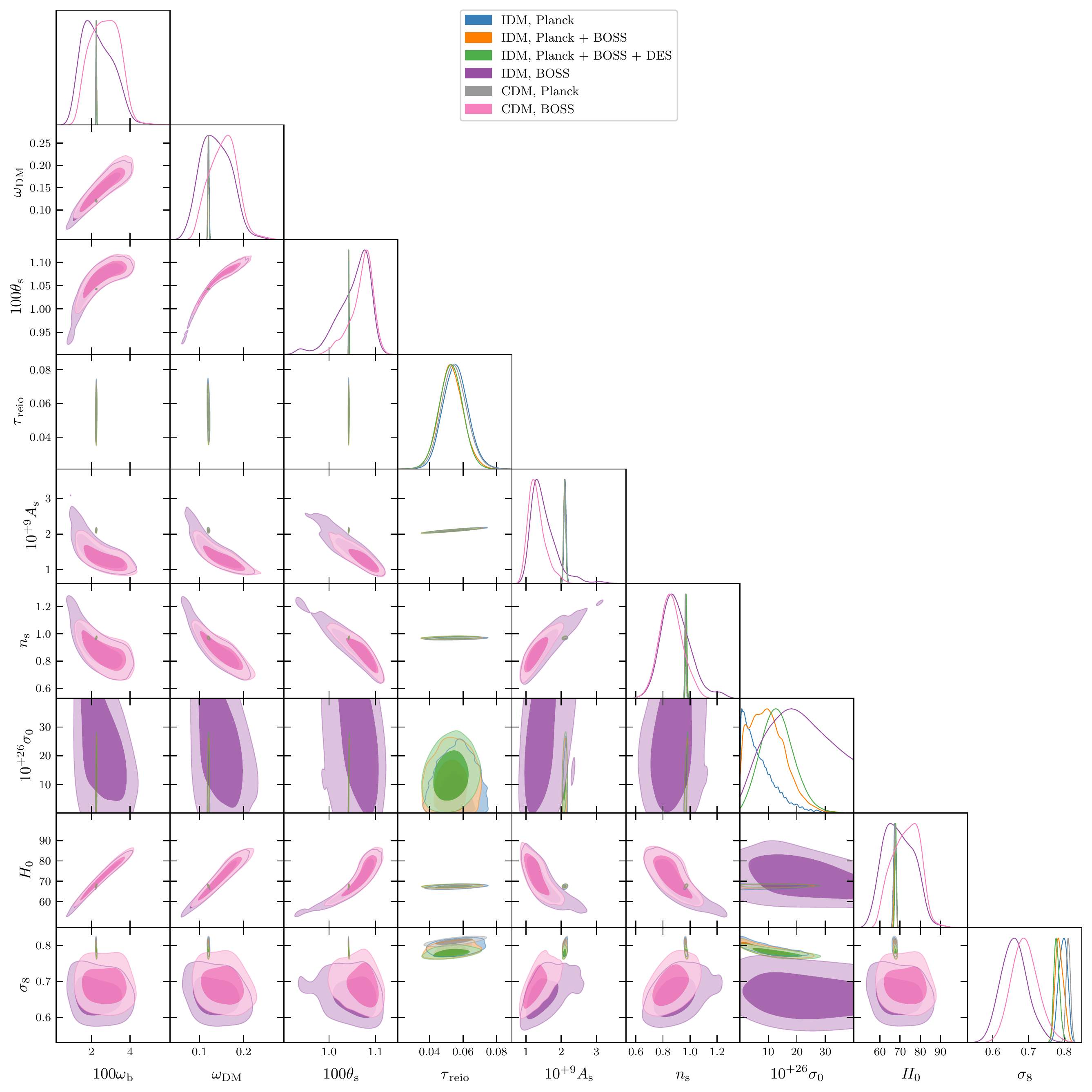}
\caption{68\% and 95\% confidence level marginalized posterior distributions of all cosmological parameters for $\Lambda$CDM from \textit{Planck} (gray) and our $m_\chi=1$ MeV, $f_\chi=10\%$ DM-baryon interacting model (colored) from different combinations of \textit{Planck}, BOSS, and DES data. Same as Figure~\ref{fig:triangle plot}, but with posteriors for BOSS-only analyses.
\label{fig:BOSS triangle plot}}
\end{figure*}

\begin{figure*}[ht!]
\includegraphics[scale=0.79]{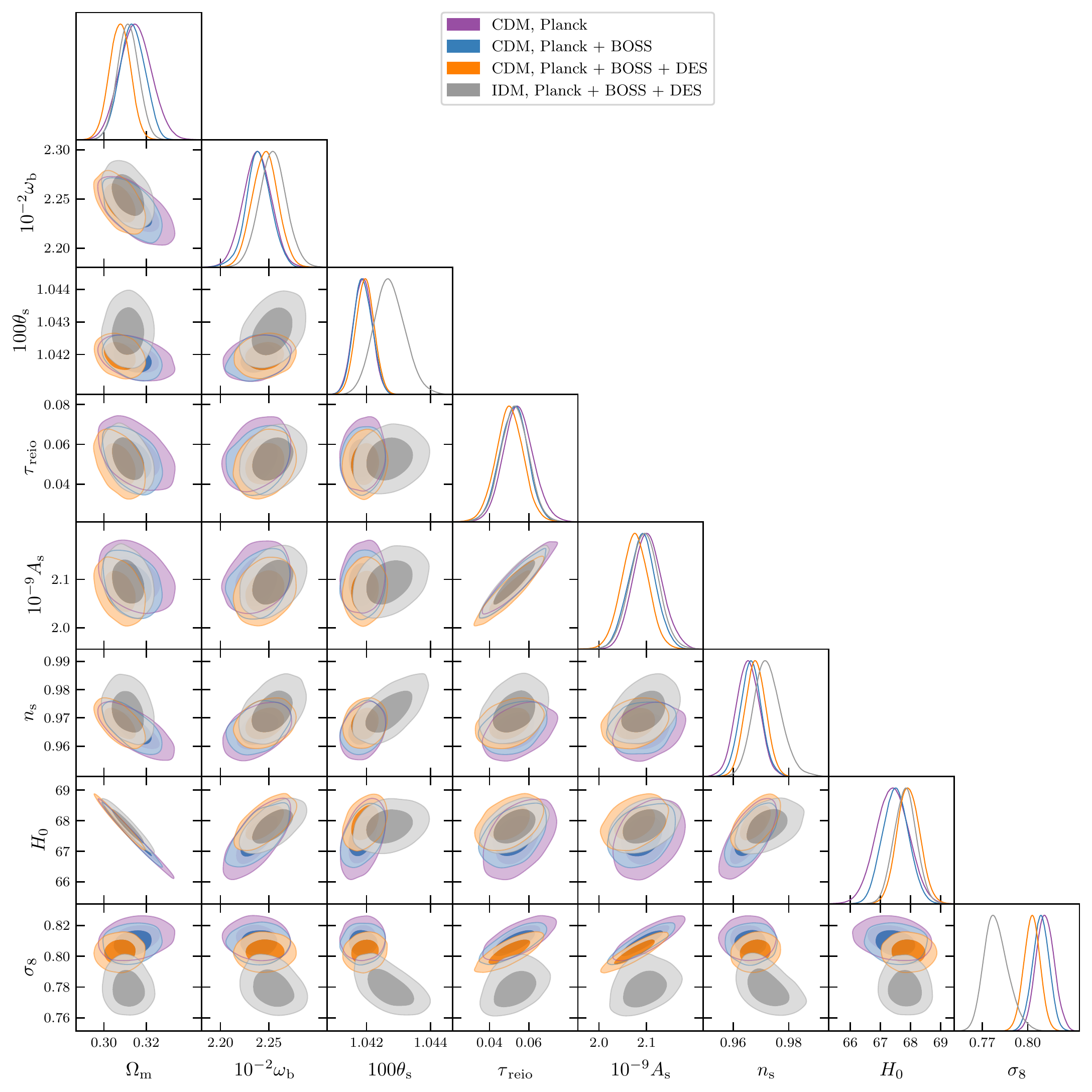}
\caption{68\% and 95\% confidence level marginalized posterior distributions of all cosmological parameters for our IDM model from a \textit{Planck} + BOSS + DES analysis (gray) and for $\Lambda$CDM (colored) from different combinations of \textit{Planck}, BOSS, and DES data.
\label{fig:CDM triangle plot}}
\end{figure*}

We show the full marginalized posterior distributions for all relevant parameters in our analysis of the $f_\chi=10\%$, $m_\chi=1$ MeV model in Figure~\ref{fig:triangle plot}. We show the same posterior distributions, along with the BOSS + DES posteriors, in Figure~\ref{fig:BOSS DES triangle plot}, and the same posteriors along with the BOSS-only posteriors in Figure~\ref{fig:BOSS triangle plot}. A CDM triangle plot with the same data combinations used in our analysis is displayed in Figure~\ref{fig:CDM triangle plot}.
\\

\clearpage

\section{Log prior results}
\label{sec:log prior}

\begin{figure*}[ht!]
\includegraphics[scale=0.65]{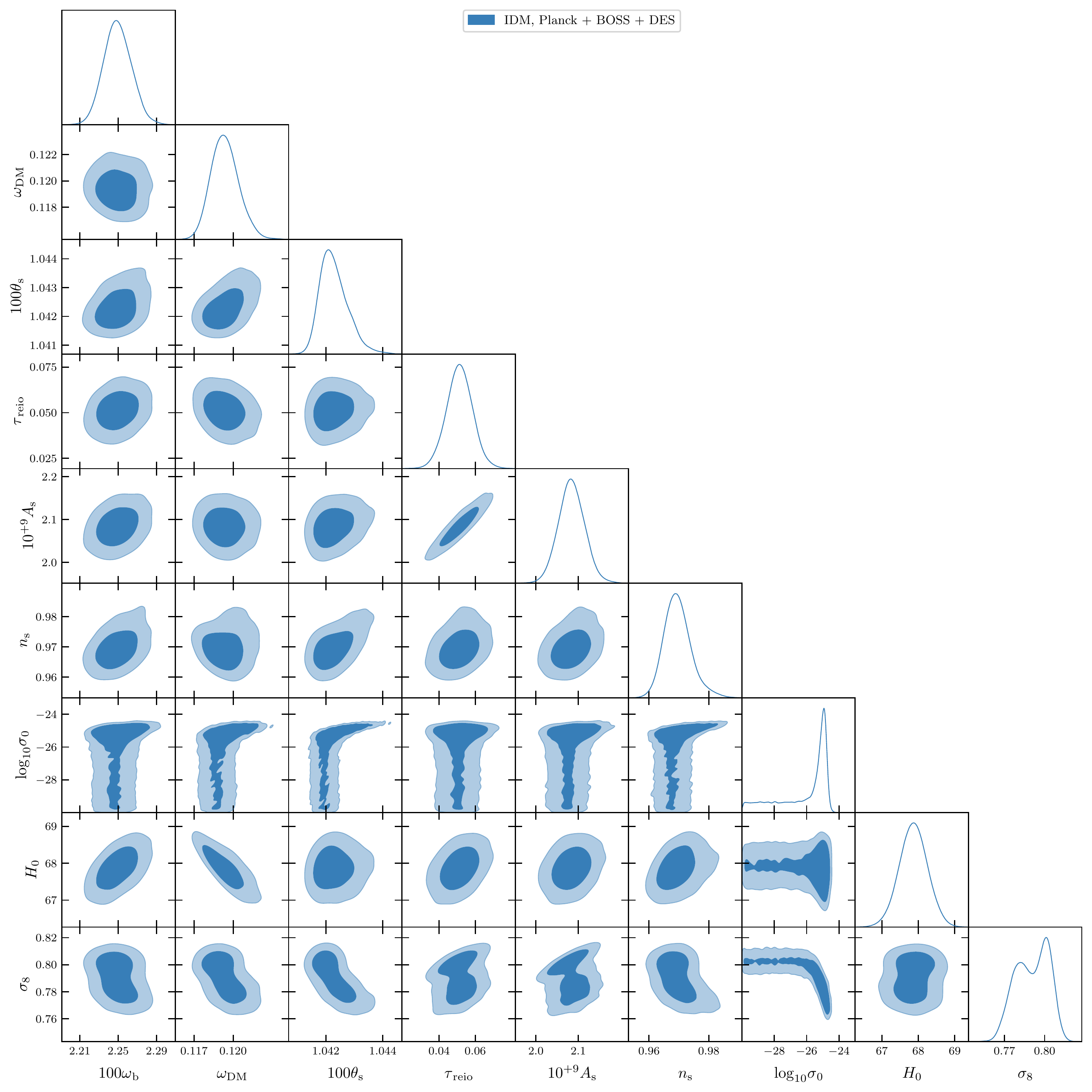}
\caption{68\% and 95\% confidence level marginalized posterior distributions of all cosmological parameters for our $m_\chi=1$ MeV, $f_\chi=10\%$ DM-baryon interacting model under a \textit{Planck}, BOSS, and DES analysis and a log prior on $\sigma_0$.
\label{fig:log}}
\end{figure*}

To assess prior dependence, we run a \textit{Planck} + BOSS + DES analysis of the $f_\chi=10\%$, $m_\chi=1$ MeV model with a log prior on $\sigma_0$. We choose the range of the log prior to be $[-30, -23]$. Table~\ref{tab:chi_log} shows marginalized limits on $\sigma_0$ and $S_8$ from this analysis, along with the $\Delta\chi^2_\mathrm{min}$ with respect to $\Lambda$CDM. We display 68\% and 95\% confidence level marginalized posterior distributions of all cosmological parameters for this analysis in Figure~\ref{fig:log}. The log prior finds $\Delta\chi^2_{\mathrm{min}}=-5.6$, which still indicates a non-negligible $>2\sigma$ preference for our IDM model over CDM. We therefore state that our IDM model shows consistent preference over $\Lambda$CDM regardless of the prior choice.

\begin{table*}[h!]
\centering
\tabletypesize{\scriptsize}
\tablewidth{0pt} 
\caption{Maximum of the marginalized posterior (maximum of the full posterior, or the best-fit value) and ± 68\% confidence level uncertainties for $\sigma_0$ and $S_8$ under a \textit{Planck} + BOSS + DES analysis of our fractional IDM model with a log prior on $\sigma_0$, compared to $\Lambda$CDM. The last row shows the difference in $\chi^2$ with respect to $\Lambda$CDM. All IDM values are for $m_\chi$ = 1 MeV, $f_\chi$ = 10\%.} \label{tab:chi_log}
\begin{tabular}{|c|c|c|c|}
\hline
Model & $\Lambda$CDM, \textit{Planck} + BOSS + DES & IDM, \textit{Planck} + BOSS + DES \\
\hline 
\hline

$\mathrm{log}_{10}(\sigma_0/\mathrm{cm}^2$)& -- & $-26.2 \ (-24.96)^{+1.6}_{-0.43}$ \\ [0.5ex]

$S_8$ & $ 0.813 \ (0.813)\pm{0.009}$ & $0.803 \ (0.795)^{+0.014}_{-0.012}$ \\ [0.5ex] 
\hline
$\Delta\chi^2_{\mathrm{min}}$ & -- & $-5.6$ \\
[0.5ex]
 \hline
\end{tabular}
\end{table*}

\clearpage

\bibliography{sample631}{}
\bibliographystyle{aasjournal}

\end{document}